# Accelerating fields up to 49 MV/m in TESLA-shape superconducting RF niobium cavities via 75°C vacuum bake

A. Grassellino,[1,2, a)] A. Romanenko,[1,2], D. Bice[1], O. Melnychuk[1], A. C. Crawford[1], S.Chandrasekaran[1], Z. Sung[1], D.A. Sergatskov[1], M. Checchin[1], S. Posen[1], M. Martinello[1], G.Wu[1]

[1]*Fermi National Accelerator Laboratory, Batavia IL, 60510, USA*

[2]*Northwestern University, Evanston, IL, 60208, USA*

In this paper we present the discovery of a new surface treatment applied to superconducting radio frequency (SRF) niobium cavities, leading to unprecedented accelerating fields of 49 MV/m in TESLA-shaped cavities, in continuous wave (CW); the corresponding peak magnetic fields are the highest ever measured in CW, about 210 mT. For TESLA-shape cavities the maximum quench field ever achieved was ~45 MV/m - reached very rarely- with most typical values being below 40 MV/m. These values are reached for niobium surfaces treated with electropolishing followed by the so called "mild bake", a 120°C vacuum bake (for 48 hours for fine grain and 24 hours for large grain surfaces). We discover that the addition during the mild bake of a step at 75°C for few hours, before the 120°C, increases systematically the quench fields up to unprecedented values of 49 MV/m. The significance of the result lays not only in the relative improvement, but in the proof that niobium surfaces can sustain and exceed CW radio frequency magnetic fields much larger than $H_{c1}$, pointing to an extrinsic nature of the current field limitations, and therefore to the potential to reach accelerating fields well beyond the current state of the art.

SRF cavities are the enabling building blocks of state of the art particle accelerators, or extremely sensitive detectors. Thousands of them are used and planned to be used in particle accelerators worldwide for various applications – light sources, particle physics, nuclear physics; their extremely high quality factors –larger than $1e^{11}$ - make these resonators unique tools even for applications like quantum computing. The performance of SRF cavities is determined by the first 100 nm of the inner cavity surface, where the supercurrents flow once the resonator is cooled below critical temperature. Recently SRF niobium resonators have seen large improvement in quality factors via nitrogen doping [1] and nitrogen infusion [2]. One of the earlier findings which brought SRF cavities to reach higher accelerating fields was the so-called 120°C bake [3], which had allowed to overcome the systematic high field Q-slope limitation. These surface treatments change (in a systematic way) the near-surface nanostructure – impurity, dislocation and vacancy content, in the first hundreds (or less) nanometers of the niobium surface. Experimental data has unequivocally shown that diffusing nitrogen in interstitial form in the surface layer leads to a dramatic improvement in quality factor, in particular via the reversal of the field dependence of the temperature dependent component of the surface resistance. However, high temperature N-doped cavities show consistently lower quench fields than with the 120°C bake surface treatment. The effect of the 120°C bake on cavity and surface nanostructure has been studied extensively [4, 5, 6], in an effort to understand the mechanisms leading to the improvement in achievable accelerating fields. It has been shown that the 120°C bake modifies only the first 60 nanometers of the surface, with the predominant theory – supported by positron annihilation data [7, 8] - being that at this temperature in niobium, vacancies start to form/diffuse, creating vacancy-hydrogen complexes [9] and preventing or reducing the formation of niobium nano-hydrides precipitates upon cavity cooldown. These nano-hydrides have been studied via TEM, AFM and several other techniques [10] and have been shown to exist in larger/smaller quantity or different phases depending on the surface treatment [11, 12]. It is therefore a plausible theory that the ultimate magnetic peak field reach of SRF niobium cavities may be currently limited by the presence of niobium nano-hydrides which are superconducting via proximity effect [13] up to a certain magnetic field, after which they become normal conducting and eventually lead to quench. With the most recent nitrogen infusion treatment [14], very high accelerating fields have been reached, consistently in the range 40-45 MV/m in 1.3 GHz TESLA-shaped cavities (170-190 mT peak surface magnetic fields). The highest field values are achieved when nitrogen is injected in the cavity at 120°C (post 3 hours at 800°C in vacuum, which dissolves fully the niobium oxide layer), while lower values are achieved if the temperature is increased in the range 140-200°C [14], pointing to a key role of the 120°C, in the presence of nitrogen, in reaching the highest fields. In preparation for nitrogen infusion studies, a fine grain cavity received from DESY lab (name 1DE3) was prepared with the standard 120°C bake, to obtain a first baseline test to compare to. This means that the cavity, after electropolishing, is baked for 48 hours in vacuum (inside the cavity) in a low temperature oven, in the presence of the surface oxide layer (differently from N infusion where the 120 bake is done in a high temperature furnace with no oxide present) . During the temperature rise to 120°C, a thermocouple stopped working, which caused the oven to go in standby mode. The cavity then lingered around 75°C for about two hours, as shown in fig.1, and few more hours around 50°C before ramping back up to 120°C. The bake cycle was then completed successfully after 48 hours at 120°C.

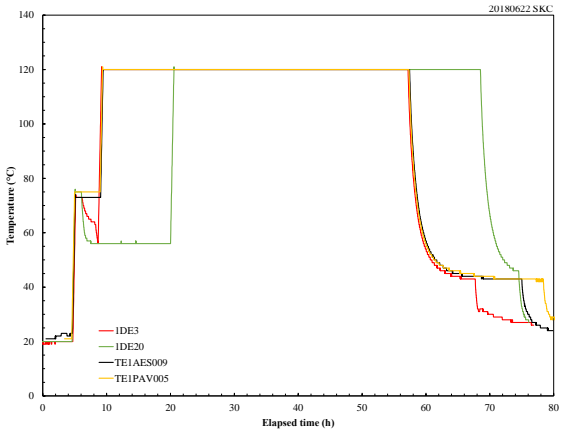

Figure 1: Low temperature vacuum bake cycle with two step process – 75°C for few hours, then 120°C for 48 hours.

The cavity received then high pressure water rinse, was assembled in clean room with low power antennas and evacuated in preparation for the vertical test where the resonator quality factor versus accelerating field is measured at T of 2K and below. Compared to the previous test the cavity had received 60 microns electropolishing and the heat treatment preparation as discussed above. In the previous test, the cavity had received EP and 120°C bake for 48 hours. As shown in fig. 2, the performance of the cavity were significantly different from previous test, for both Q and accelerating field which went from 37 to 49 MV/m. This accelerating field is unprecedented for TESLA-shape cavities, and the highest CW peak magnetic field ever achieved for niobium cavities, of ~210 mT. The Q was also improved, with $Q>10^{10}$ up to the maximum field.

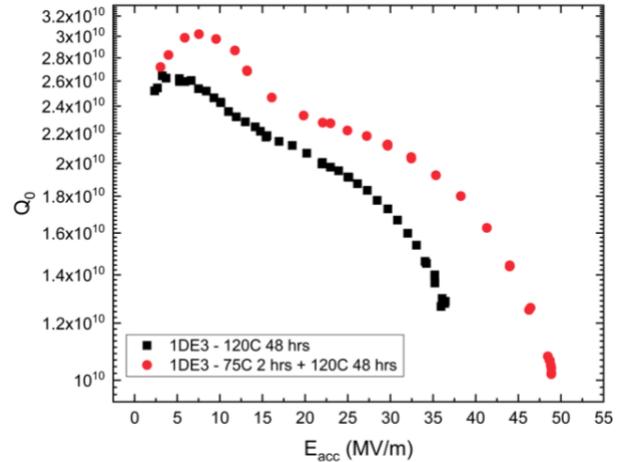

Figure 2: Comparison for same cavity 1DE3 for EP+ regular 120°C bake, and after 60 microns EP followed by 75+120°C.

While still in within the boundaries of the theoretical superheating field of niobium [15], the result was very surprising. This triggered first a re-test after several calibrations of the measurement system electronics, which confirmed outstanding results. We then performed a review of the cavity surface treatment history, and the anomaly described above during the low temperature bake was noted. The treatment with the same anomaly was repeated on a second fine grain cavity, fabricated in the US, TE1AES009. The cavity received 60 microns EP and a "two step vacuum bake" with this time 4 hours at 75°C and 48 hours at 120°C, as shown in fig. 1. To a great surprise we found that the extraordinary results were reproduced – 49 MV/m with higher Q above $10^{10}$ up to the quench field at 2K, as shown in fig. 3. This pointed to a real effect of the 75°C step on cavity performance. A literature search on niobium and the effect of low temperature bake pointed to few papers describing changes occurring in niobium in terms of vacancy content at around 70°C, in addition to changes occurring in the range ~120°C [16]. Positron annihilation data [7] had shown indeed changes in niobium cavity cutouts occurring in the range 70-140°C. One paper suggests that changes in niobium at 70°C involve vacancies while at 120°C involve dislocations (Bordoni and Hasiguti peaks) [].

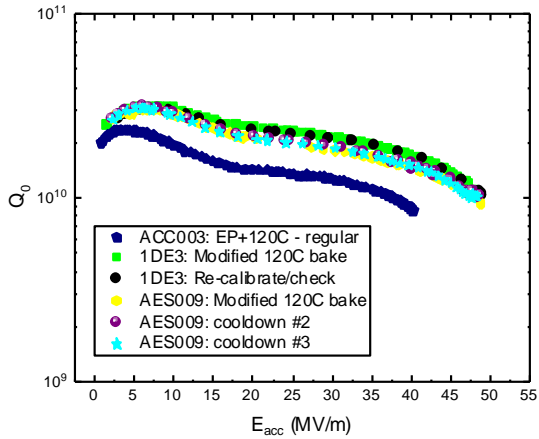

Figure 3: Comparison of 1DE3 for regular 120°C bake and for 75°C+120°C bake.

On an intuitive level, one possible explanation to the observed improved performance is the suppression of nano-hydrides via defect trapping, with defects-H complexes forming at the two different temperatures. Formation of nano-hydrides for 120°C only baked samples versus 75+120°C is under study and will be subject of future publications.

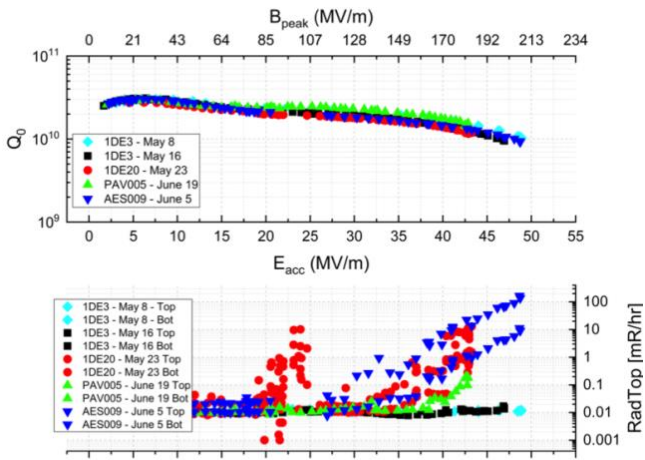

Figure 4: Comparison of all four cavities treated with 75/120°C bake. Upper Q vs E curves, lower corresponding x-rays detected during the measurements.

Fig. 4 shows the results of two additional cavities that have received the 75+120°C, one a large grain cavity 1DE20 and a fine grain cavity TE1PAV005. We can see that the all four cavities show very systematic Q vs E behavior, and the gradient achieved for the other two was 43 MV/m (190 mT) limited though by the extrinsic effect of field emission. The Q curves were measured at 2K and at 1.4K, to allow for decomposition of the surface resistance into the temperature dependent and temperature independent parts [17]. While the field dependence of the residual resistance was found to be the same as per 120°C bake cavities, the temperature dependent component was found to be lower than for the 120°C bake cavities, and actually very similar to that of nitrogen infused cavities at 120°C, as shown in fig. 5. This systematic change brought by the addition of the 75°C step in the field dependence of the temperature dependent part of the surface resistance is responsible for the systematic improvement in Q. It is also possibly indicative that the subtle nano-structural changes occurring in niobium at 75°C are related to suppression of the proximity coupled nano-hydrides and therefore a better temperature dependent surface resistance.

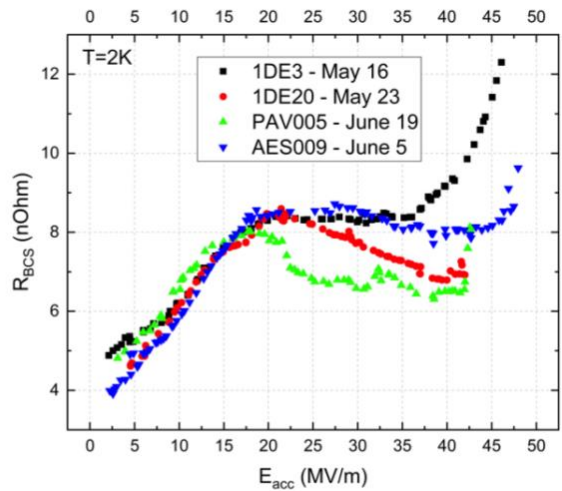

Figure 5: Comparison of 1DE3 for regular 120°C bake and for 75°C+120°C bake.

## ACKNOWLEDGMENTS

We would like to acknowledge our SRF colleagues from DESY for providing two of the cavities on which these studies were performed, and for suggesting to test the cavity to establish a baseline test, which lead to the incidental first findings. We thank Sergey Belomestnykh for fruitful discussions on this work. The work was supported by the DOE Office of High Energy Physics, via the GARD program. Fermilab is operated by Fermi Research Alliance, LLC under Contract No. DE-AC02-07CH11359 with the United States Department of Energy.